**lane level joint control of off-ramp and main line speed guidance on expressway in rainy weather**

**Boyao Peng**
School of Transportation Engineering
Chang'an University, Xi'an, China, 710000
2021903165@chd.edu.cn

**Lexing Zhang**
School of Transportation Engineering
Chang'an University, Xi'an, China, 710000
2021902169@chd.edu.cn

**Enkai Li**
Expressway College
Chang'an University, Xi'an, China, 710000
2021901597@chd.edu.cn

Word Count: 6,238 words + 4 table (250 words per table) = 7,238 words

*Submitted [31th July, 2023]*

*Boyao Peng, Lexing Zhang and Enkai Li*

**ABSTRACT**

In the upstream of the exit ramp of the expressway, the speed limit difference leads to a significant deceleration of the vehicle in the area adjacent to the off-ramp. The friction coefficient of the road surface decreases under rainy weather, and the above deceleration process can easily lead to sideslip and rollover of the vehicle. Dynamic speed guidance is an effective way to improve the status quo. Currently, there is an emerging trend to utilize I2V technology and high-precision map technology for lane-level speed guidance control. This paper presents an optimized joint control strategy for main line-off-ramp speed guidance, which can adjust the guidance speed in real time according to the rainfall intensity. At the same time, this paper designs a progressive deceleration strategy, which works together with the speed guidance control to ensure the safe deceleration of vehicles. The simulation results show that the proposed control strategy outperforms the fixed speed limit control in terms of improving the total traveled time (TTT), total traveled distance (TTD) and standard deviation of speed (SD). Sensitivity analysis shows that the proposed control strategy can improve performance with the increase of the compliance rate of drivers. The speed guidance control method established in this paper can improve the vehicle operation efficiency in the off-ramp area of the expressway and reduce the speed difference of each vehicle in rainy weather, which guarantee the safety of expressway driving in the rainy day.

## INTRODUCTION

Statistics from the National Expressway Traffic Safety Administration (NHTSA) stated that rain causes 46% of weather-related car accidents (*1*). Rainfall has become one of the major causes of expressway crashes. In rainy weather, the driving distance is shortened and the adhesion coefficient of pavement is reduced, the vehicle is very easy to skid out of control. Compared to the main line of the expressway, the accident rate of ramps is about 1/3 higher in rainy weather. The reason is that ramps usually have narrower lane widths, more curved alignments, and greater longitudinal slopes (*2-3*). As a result, ramps also typically have lower safety speed limits. Whereas the high speed limit on the main line of the expressway varies greatly from the speed limit on the ramps. This results in the majority of vehicles still entering the off-ramp at higher speeds in rainy weather, which increasing the risk of accidents. Z. Song designed a control system for progressive deceleration of off-ramps and verified through simulation that it can effectively avoid vehicle accidents (*4*).

Progressive deceleration control belongs to the variable Speed Limit (VSL), which is a control method that intervenes in the operation of traffic flow by adjusting the speed limit according to certain strategies (*5-8*). Carlson. R. C. et al. (*9*). Regulated the speed limit value of the upstream section of the bottleneck, which avoided the reduction of capacity at the bottleneck and improves the efficiency of access. However, the current VSL control mainly relies on Variable Message Sign (VMS) to assign a uniform speed limit to all lanes of a certain road segment and therefore is not effective enough (*10-11*). Studies have shown that expressways have a distinction between fast and slow lanes, and that there are differences in traffic conditions between different lanes at the same time (*12*). Letter demonstrates that variable speed limit control with different speed limits between lanes has better results compared to VSL using micro-simulation based on data from a real scenario (*13*).

In recent years, there has been an explosion in high-precision mapping technology, which provides an accurate representation of road geometry, on-road facilities, and the ability to provide lane level guidance to vehicles (*14*). Currently, companies such as Baidu and Google are already conducting road trials on high-precision maps (*15-16*). With the support of high-precision maps, VSL will be further developed into lane level single-vehicle speed guidance control. T. Zhang proposed a novel lane level road network model, which guarantee the global continuity for arbitrary map route and better approximates the real vehicle trajectory (*17*). Furthermore, in the case of poor visibility in rainy weather, the rich road geometry details of high-precision maps can give drivers the ability to perceive over-the-horizon (*18*). In this paper, we propose a joint control strategy for lane level speed guidance of main line and off-ramp based on I2V environments supported by high-precision maps.

Macroscopic traffic flow model METANET for Expressways is Used to Construct Speed Guidance Algorithms in this paper. The METANET is widely used in constructing VSL control algorithms (*19-21*). However, the classical form of the model does not have the accuracy of lane level control and does not take into account the effect of rainfall on the traffic flow. Therefore, in this paper, the METANET is further discretized to the lane level. L. Han proposed a lane level METANET model that adds virtual off-ramp lane to, which corresponding the number of traffic flows and the number of lanes in different directions (*22*). However, the speeds of straight and exiting traffic can affect each other, The method of adding virtual lanes ignores this consideration. Hence, the effect of PDS on traffic flow is taken into account and the METANET model is modified.

The main contributions of this paper are as follows:

1. The progressive deceleration strategy in upstream off-ramp were proposed, which ensures that the off-ramp vehicles decelerate smoothly to the off-ramp.
2. In the control framework, the effect of safe speed in rainy weather and the joint speed guidance control of main line and off-ramp were considered, which are not considered in the traffic flow model with VSL control.
3. The simulation results of various scenarios were compared to validate the effectiveness and generalization of the proposed method.

**METHODS**

**Metanet Model**

The METANET model uses Taylor's formula and differential equations to dwascretize the traffic flow parameters in time and space. Which recurses the state of the adjacent sgments at the next discrete time based on the traffic parameters of the current segment at the current time. Traffic flow on expressway is continuous because it does not contain level crossings. Therefore, the traffic flow was divided into segments in discretization process. For ease of description, the expressway was divided into $M$ segments, each of length $x_i$ (m) (**Figure 1**). Time was dcretized into sampling periods of equal length $\Delta t$. The traffic flow evolution model is as follows:

$$q_i(\lambda) = k_i(\lambda)v_i(\lambda), \qquad (1)$$

$$k_i(\lambda + 1) = k_i(\lambda) + \frac{\Delta t}{x_i}[q_{i-1}(\lambda) - q_i(\lambda) + r_i(\lambda)], \qquad (2)$$

$$v_i(\lambda + 1) = v_i(\lambda) + \frac{\Delta t}{\tau}[V_h[k_i(\lambda)] - v_i(\lambda)] + \frac{\Delta t}{x_i}v_i(\lambda)[v_{i-1}(\lambda) - v_i(\lambda)] - \frac{1}{\tau}\left[\frac{\eta \Delta t}{x_i} \cdot \frac{k_{i+1}(\lambda) - k_i(\lambda)}{k_i(\lambda) + \kappa}\right], \qquad (3)$$

where $q_i(\lambda)$ = traffic volume of segment *I* during discrete time $\lambda$ *(veh/h)*; $k_i(\lambda)$ = traffic density of segment *I* during discrete time $\lambda$ *(veh/km)*; $v_i(\lambda)$ = space mean speed of segment *I* during discrete time $\lambda$ *(km/h)*; $r_i(\lambda)$ = on-ramp volume of segment *I* during discrete time $\lambda$ *(veh/h)*; $x_i$ = length of segment *I (m)*; $\tau$ = driver's delay adjustment factor, $\eta$ = velocity-density relationship coefficient, $\kappa$ = Elasticity factor, all three are calibrated by road traffic flow data; $V_h[k_i(\lambda)]$ = desired speed at density $k_i(\lambda)$, which is calculated by the following equation:

$$V_h[k_i(\lambda)] = \min\left\{v_{i,j}^f \cdot exp\left[-\frac{1}{a}\left(\frac{k_i(\lambda)}{k_i^c}\right)^a\right]\right\}, \qquad (4)$$

where $v_{i,j}^f$ = free-flow speed of lane *j* at segment *I (km/h)*; $k_i^c$ = critical density of segment *I (veh/km)*; $a$ = model parameters, calibrated by velocity-density diagram.

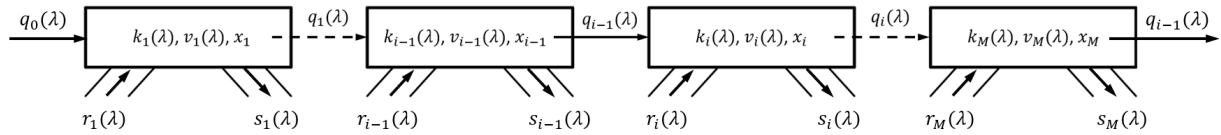

**Figure 1 Schematic diagram of segments on expressway**

**Progressive Deceleration Control**

The main line of the expressway typically has three lanes, numbered from left to right of the direction of travel, joins the on-ramp and off-ramp on the right. Vehicles leaving the expressway should move to the right-hand lane (lane 3) before the exit. Due to the large difference in speed limits between the main line and the ramp, vehicles tend to maintain high speeds and brake significantly when entering the off-ramp. During rainfall, the speed of the vehicle exceeds the skidding speed threshold and the vehicle is highly susceptible to accidents. Therefore, it is necessary to give the exiting vehicle a sufficiently gentle deceleration by means of a progressive deceleration. Figure 2 illustrates the progressive deceleration segment (PDS) proposed in this paper, which is located in lane 3 upstream of the off-ramp. In the vehicle-road collaboration scenario supported by high-precision maps, when the vehicles enter the PDS, the road equipment will send real-time safe speed recommendations to the drivers to ensure that the vehicle arrives at the off-ramp at the safe speed. Assuming that the vehicle decelerates within the PDS at a constant deceleration rate, the length of the PDS $l_d$ (m) can be calculated with the following equation:

$$l_d = \frac{V_r^2 - V_{j_3}^2}{2a_o}, \qquad (5)$$

Where $V_r$ = safe speed limit of off-ramp *(km/h)*. $V_{j_3}$ = safe speed for lane 3 *(km/h)*; $a_o$ = optimal deceleration of vehicles under specific rainfall intensity *(m/s²)*. Obviously, safety deceleration is the key to calculating the desired speed for the PDS. A detailed study about pavement conditions shows the relation between friction force and vehicle speed by levels of water film depth *(23)*. Combining the relation in the study with pavement material information, safe deceleration is obtained by correlating to the pavement friction coefficient. This method gives the maximum safe deceleration $a_{max}$. However, in reality, the pavement friction coefficient is dynamic and therefore there is a range of values for $a_o$. In this paper, $a_0$ is an optimization variable to be determined by solving the optimization equation.

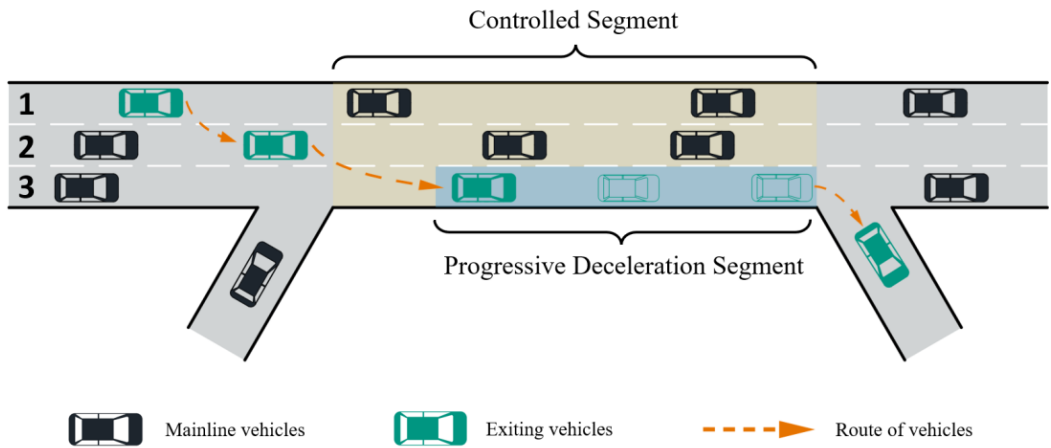

**Figure 2 Schematic diagram of progressive deceleration segment (PDS) proposed in this paper**

## METANET For Joint Control of Lane Level Speed Guidance

To develop METANET model for lane level joint control of off-ramp and main line speed guidance with PDS, the following traffic characteristics were considered before modeling the METANET:

1.  The model developed in this paper is based on the vehicle-road collaboration environment, assuming that all vehicles support I2V communication and are equipped with lane level high-precision map guidance software.
2.  Typically, vehicles exiting the expressway will change lanes to the right lane upstream of the impact area of the exit ramp. As a result, the roadside I2V device will issue a lane change alert upstream of the control area to vehicles intending to exit. It is assumed in this model that all exiting vehicles are already in lane 3 when they enter the control area.
3.  Different vehicles have different geometries and driving characteristics, as well as different stopping distances and tire skid resistance, which need to be considered differently in the calculation of variable speed limits. To simplify the model, this paper assumes that all vehicles are small cars.

Firstly, the metanet model is further discretized to lane level. Use *N* to indicate the number of lanes, numbered from the left to the right of the direction of travel. Secondly, considering that the METANET model is applied to expressways, each section is long, and each section in Kotsialos Papageorgiou's paper is around 10 *km (24)*. However, in the area upstream of the exit ramp, where vehicles frequently adjust their travel status, especially under the PDS proposed in this paper, the effect of the upstream speed represented by the 3rd term on the right side of the middle sign in Eq. (3) is amplified exponentially, and, therefore, a coefficient $\omega$ is added to this term to regulate the effect of this component. Thirdly, Under the joint control of speed guidance, the traffic flow condition of the whole expressway will no longer be homogeneous, so the traffic flow parameters of the speed guidance control segment will no longer be able to be directly derived from the parameters of the previous segment, and this effect can be reflected in the speed-density sensitivity parameter $\eta$, using the following formula to implement the correction of equation (3):

$$\eta(\lambda) = -v_{i,j}^g \cdot \frac{[k_{i,j}(\lambda)]^{a-1}}{(k_{cr}^d)^a} \cdot e^{-\frac{[k_{i,j}(\lambda)]^a}{a \cdot k_{cr}^d}}, \qquad (6)$$

Where $v_{i,j}^g =$ guidance speed speed under lane level joint control of speed guidance of lane $j$ at segment $I$; $k_{cr}^d =$ critical density under lane level joint control of speed guidance.

Hence, to meet the requirements of lane level joint control of off-ramp and main line speed with PDS, the METANET model needs the following modifications:

$$q_{i,j}(\lambda) = k_{i,j}(\lambda) v_{i,j}(\lambda), \qquad (7)$$

$$k_{i,j}(\lambda+1) = k_{i,j}(\lambda) + \frac{\Delta t}{x_i}[q_{i-1,j}(\lambda) - q_{i,j}(\lambda) + r_{i,j}(\lambda)], \qquad (8)$$

$$v_{i,j}(\lambda+1) = v_{i,j}(\lambda) + \frac{\Delta t}{\tau}[V_h[k_{i,j}(\lambda)] - v_{i,j}(\lambda)] + \omega \frac{\Delta t}{x_i} v_{i,j}(\lambda)[v_{i-1,j}(\lambda) - v_{i,j}(\lambda)] - \frac{1}{\tau}\left[\frac{\eta(\lambda)\Delta t k_{i+1,j}(\lambda) - k_{i,j}(\lambda)}{x_i k_{i,j}(\lambda) + \kappa}\right], \qquad (9)$$

where $q_{i,j}(\lambda) =$ traffic volume of lane $j$ at segment $I$ during discrete time $\lambda$ *(veh/h)*; $v_{i,j}(\lambda) =$ space mean speed of lane $j$ at segment $I$ during discrete time $\lambda$ *(km/h)*; $k_{i,j}(\lambda) =$ traffic density of lane $j$ at segment $I$ during discrete time $\lambda$ *(veh/km)*; $r_{i,j}(\lambda) =$ on-ramp volume of lane $j$ at segment $I$ during discrete time $\lambda$ *(veh/h)*; $\omega =$ adjustment coefficient for upstream speed effects.

With the lane level joint control of speed guidance, the desired speed $V_h[k_{i,j}(\lambda)]$ is the smaller of the guidance speed $v_{i,j}^g$ and the variable speed limit estimated from the density. In addition, a speed-density model for rainy weather developed by previous studies (use the function $F[k_{i,j}(\lambda), L_v]$ to represent) has been used to characterize the impact of visibility on traffic flow, which is expressed by the following equation:

$$F[k_{i,j}(\lambda), L_v] = A \cdot e^{(B \cdot k_{i,j}(\lambda) + C \cdot L_v)}, \qquad (10)$$

Where $L_v =$ visibility; $A$, $B$ and $C$ are constants, calibrated from measured data on rainy weather.

Hence, $V_h[k_{i,j}(\lambda)]$ can be modified as:

$$V_h[k_{i,j}(\lambda)] = \min\left\{v_{i,j}^f \cdot exp\left[-\frac{1}{h}\left(\frac{k_{i,j}(\lambda)}{k_{i,j}^c}\right)^h\right], (1+\gamma)v_{i,j}^g(\lambda), F[k_{i,j}(\lambda), L_v]\right\}, \qquad (11)$$

Where $k_{i,j}^c =$ critical density of lane $j$ at segment $I$ *(veh/km)*; $\gamma =$ rate of non-compliance with speed limits by drivers considering the actual situation *(%)*; $h =$ a parameter characterizing the impact of rainfall on traffic flow, which is a modification applied to the velocity-density equation used in the classical form of the METANET. A proposed velocity-density model under rainfall *(25)*. indicates $h$ range of roughly 1.3 to 2, $h$ takes the value 2 in this paper.

In addition, guidance speed for lane 3 with progressive deceleration control is as follows:

$$v_{i,3}^g[\lambda, l(\lambda)] = \sqrt{2l(\lambda)a_o + V_{j_3}^2}, \qquad (12)$$

Where $l(\lambda) =$ distance of the vehicle from the start of the PDS at discrete time $\lambda$ *(m)*, the starting position of deceleration is calculated based on the length of the PDS $l_d$ *(m)*.

**Objective Function and Constraints**

In this paper, travel safety and traffic efficiency are considered as the optimization objectives of the multi-objective function. The minimum total traveled time (TTT) and the maximum total traveled distance (TTD) are taken as the control objectives for traffic efficiency. Standard deviation of speed (SD) is taken as the control objectives for travel safety. In addition, a total travel time weighting factor $\alpha_{TTT}$, a total distance weighting factor $\alpha_{TTD}$ and a weighting factor $\alpha_{SD}$ for standard deviation of speed are introduced. The objective function Is as follows:

$$J = \Delta t \sum_{\lambda=1}^{n_T} \sum_{i=1}^{M} \sum_{j=1}^{N} x_i \cdot \left[\alpha_{TTT} k_{i,j}(\lambda) - \alpha_{TTD} k_{i,j}(\lambda) v_{i,j}(\lambda)\right] + \alpha_{SD} SD, \qquad (13)$$

Where $n_T =$ number of discrete-time iterations in the control period, which is equal to the ratio of $\Delta t$ to the control period $T$. $SD$ reflects traffic conflicts and can indirectly express the level of safety of expressway operations, which can be obtained from real-time detection data. For each lane, $SD_j$ is as follows:

$$SD_j = \sqrt{\frac{1}{4} \sum_{i=1}^{M} \frac{[\bar{v}_{i,j} - v_n]^2}{q_{i,j}}}, \qquad (14)$$

Where $v_n =$ the speed of the n$^{\text{th}}$ car; $\bar{v}_{i,j} =$ Average speed of all vehicles in lane $j$ on segment $i$.

The variable guidance speed should be set with the safety of driving under rainy weather, so set the following constraints:

1. The guidance speed for segment $I$ must not exceed the maximum safe speed $V_i^{max}$ (the calculation method is described below) for the current rainfall intensity or the legal speed $V_i^{limit}$ limit on segment $i$:

$$V_{i,j}^g(\lambda) \leq \min\left(V_i^{max}, V_i^{limit}\right), \qquad (15)$$

2. When the road is wet and slippery in the rain, sharp speed changes on the road are undesirable. Therefore, the speed difference between adjacent segments should not be greater than 20 km/h (*26*):

$$\left|V_{i+1,j}^g(\lambda) - V_{i,j}^g(\lambda)\right| \leq 20 \qquad (16)$$

3. To avoid shocks to driver behaviour and traffic flow stability caused by rapid changes in speed limit values, the difference in speed limits between adjacent cycles $T_k$ and $T_{k+1}$ on the same road segment should not be greater than 20 km/h:

$$\left|V_{i,j}^g(T_{k+1}) - V_{i,j}^g(T_{k+1})\right| \leq 20.0, \qquad (17)$$

4. To ensure safe deceleration, the optimum deceleration of the progressive deceleration segment in each control cycle must not be greater than the maximum safe deceleration $a_{max}$:

$$a_o(\Delta t_k) \leq a_{max}, \qquad (18)$$

**Safe Speed on Off-Ramp**

Safe speed on the off-ramp is the target for progressive deceleration of vehicles on the off-ramp and is therefore key to joint control of speed guidance. **Figure 3(a)** shows the coordinate system for a vehicle travelling on pavement. On ramps, the lateral and sideways forces on the vehicle are dominant and the forcea are analysed as in **Figure 3(b)**.

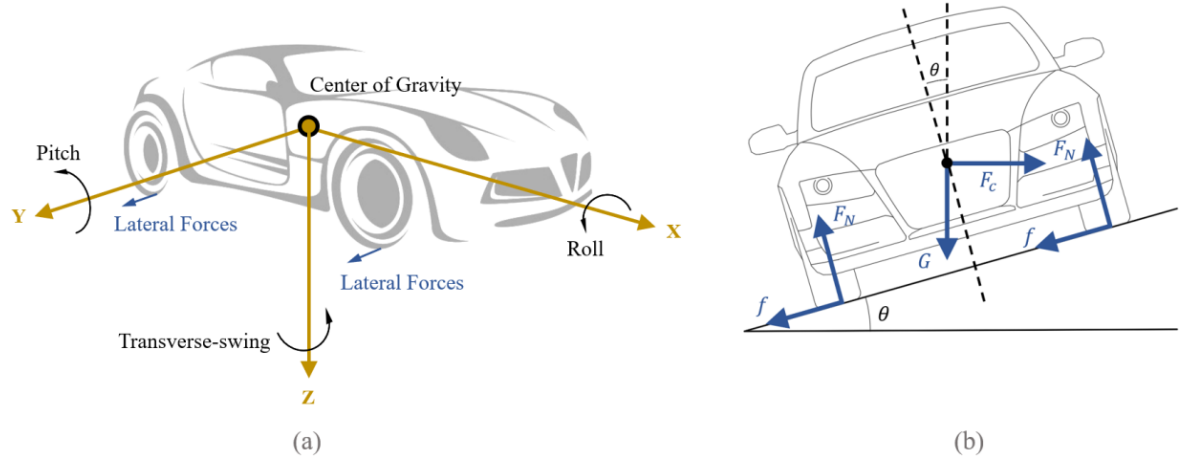

(a)                                                                                      (b)

**Figure 3 coordinate system for car: (a) Mechanical analysis of vehicles on ramps: (b)**

Where $G =$ gravity *(F)*; $F_c =$ centripetal force *(F)*; $f =$ lateral friction resistance of tyres *(F)*; $F_N =$ supporting forces on tyres *(F)*; $\theta =$ superelevation of pavement *(°)*.

When the friction between the vehicle and the pavement is not sufficient to provide the required centripetal force, resulting in a tendency for the vehicle to slide outwards. Further, it can be considered that the vehicle skids when the lateral wheel attachment force is less than the lateral frictional resistance between the pavement and the tyres. Moreover, the lateral frictional resistance is related to the coefficient of road adhesion. Some effective methods have been proposed for calculating the coefficient of road adhesion as the following equation (*27-28*):

$$\varphi = 0.8256 - 0.0043v_{max} - 0.0072h, \qquad (19)$$
$$h = 0.1258 \times l^{0.6715} \times i^{-0.3147} \times d^{0.7786} \times TD^{0.7261}, \qquad (20)$$

Where, $\varphi =$ coefficient of road adhesion; $h =$ water film depth on pavement *(mm)*; $l =$ slope length of off-ramp *(m)*; $i =$ longitudinal gradient of the off-ramp *(%)*; $d =$ intensity of rainfall *(mm/min)*; $TD =$ depth of pavement structure *(mm)*.

In addition, the safety speed limit is related to the line of the off-ramp, the smaller the radius of the curve, the greater the centrifugal force on the vehicle, the more likely to exceed the friction between the vehicle and the pavement. In this paper, carsim is used in the dynamics simulation to collect the critical speed of the vehicle under different coefficient of road adhesion and curve radius of off-ramps. The quantized model is obtained by surface fitting as follows:

$$v_{max} = 0.782R + (68.457 + 0.247R)\varphi - 0.00335R^2 - 32.171\varphi^2 - 5.272, \qquad (21)$$

Where $R =$ curve radius of off-ramps.

**Safe Speed Limit on Main Line**

Safe speed $V_i^{max}$ on the main line were mentioned earlier, and since the gradient and alignment of the main line are generally excellent, its main consideration is SSD in rainy weather. The vehicle braking and deceleration process is shown in **Figure 4**. Where $L_v =$ visibility *(m)*; $L_1 =$ distance traveled by the vehicle in front *(m)*; $L_2 =$ distance traveled by the rear driver in reaction time *(m)*; $L_3 =$ distance driven from braking to stopping of rear vehicle *(m)*; $L_s =$ Safe distance between front and rear vehicles *(m)*, generally taken as 5*m*;

Consider the most unfavorable case, where the obstacle ahead is at rest in the case of traffic accidents, scattered goods, etc. $l_1 = 0$. At this time, the rear vehicle on the other side can stop safely under the condition that:

$$l_2 + l_3 + l_s \leq L_v, \qquad (22)$$
$$l_2 = v_{i,j} \cdot t_r, \qquad (23)$$
$$l_3 = \frac{v_{i,j}}{2 \cdot 9.8 \cdot \varphi}, \qquad (24)$$

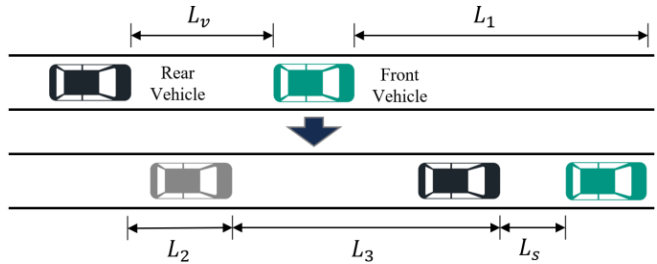

**Figure 4 Diagram of a vehicle slowing down to a stop**

In the above equation, $t_r =$ driver's reaction time. The the relationship between visibility and rainfall intensity is as follows：

$$L_v = 294.8d^{-1.1}, \qquad (25)$$

Substituting Eqs. (19), (20) and (23) into Eq. (22) yields the safe speed limit for the main line as follows:

$$v_{max} = (0.224h - 0.804d^{-1.1} - 25.63) + 0.634 \times [(0.353h - 1.268d^{-1.1} - 40.43)^2 - 3.156 \times (294.8d^{-1.1} - 5) \times (0.0072h - 0.826)]^{\frac{1}{2}}, \qquad (26)$$

**Framework of Joint Control for Speed Guidance**

As described in the previous section, the function of the PDS is to smooth the deceleration from main line to off-ramp in rainy weather, which is closely related to the safe speed of main line and off-ramp. In rainy weather, visibility and adhesion coefficients of pabement are various for different rainfall intensities, and therefore safe speed limits are also different. This paper proposes a new framework for joint control of variable speed guidance of main line and off-ramp on expressway. The framework is shown in **Figure 5**. In each control cycle $T_k$, the vehicle detectors collect the traffic state and inputs it into the METANET model to predict the future traffic state, which requires cyclic calculation at discrete time intervals until a cycle length of $T_0$; Secondly, the traffic parameters are passed to the objective function and constraints; at the same time, the rain gauge measures the real-time rain intensity to compute the safe speed, which is passed to the constraints; Finally, the guidance speed $v_{i,j}^g$ obtained by the algorithm will be averaged and released to the vehicle through I2V system.

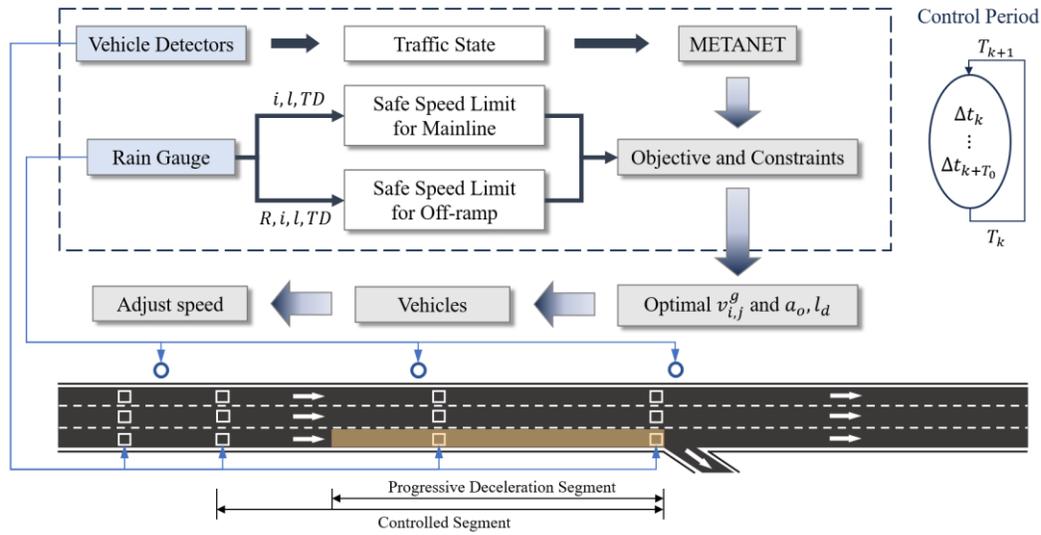

**Figure 5 Framework of joint control for speed guidance**

## Simulation Setup

A model of cellular automata (CA) developed based on the mathematical software MATLAB R2022a was used to simulate the experiments. Vehicle operation in CA follows the improved NaSch rule; The size of the experimental vehicle was *4.3m × 1.8m (L × W)*; The size of the cell was *0.5m × 0.5m*; The control period is 5*min*, the discrete time $T_0$ is 20*s*, and the simulation duration is 60*min*. A 2-km real-world east bound asphalt-surfaced motway section with off-ramp was selected in Xi'an, China. Vehicle detectors were installed on gantries along the main line of this section and were used to capture traffic volumes, vehicle speeds, and traffic densities. This section was selected because of the simplicity of its main line shape, as well as the number and width of lanes in accordance with the modeling design in this paper. **Figure 6(a)** shows a map of the study section downloaded from Google Maps; **Figure 6(b)** shows the extracted geometric lineaments of this road section, which were used to construct the CA model.

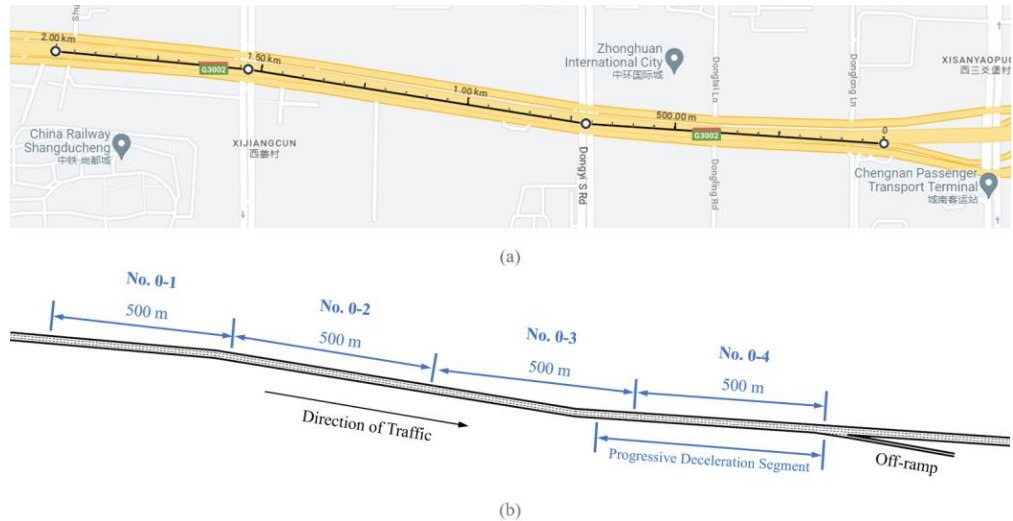

**Figure 6 Expressway sections in GoogleMap: (a) Segmentation of road sections: (b)**

Before simulation, the model parameters first need to be calibrated. In order to improve the accuracy of model applications, model parameters are generally calibrated using real or simulated data. A method for calibrating model parasmeters based on parameter estimation was applied to the quantitative model validation procedure designed by Kokkalis and Panagouli (*24*). Data from video detectors on the study section for three consecutive days (April 6–8, 2023) were available. **Table 1** shows the basic parameters of each segment. In this paper, the nonlinear fitting tool of MATLAB is used to calibrate the model parameters by combining the above data and the least squares method, and the parameters of METANET model after calibration of the above data are shown in **Table 2**.

**TABLE 1 Basic parameters of each segment**

| Segment No. | Lane No. | Free flow speed $v_f$ / km/h | Critical density $k_c$ / veh/km | capacity $C_p$ / veh/h |
|:---:|:---:|:---:|:---:|:---:|
| | 1 | 124.3 | 328 | 435 |
| 0-1 | 2 | 105.7 | 286 | 239 |
| | 3 | 87.7 | 228 | 143 |
| | 1 | 120.6 | 483 | 402 |
| 0-2 | 2 | 100.2 | 286 | 249 |
| | 3 | 80.5 | 196 | 164 |
| | 1 | 114.6 | 481 | 395 |
| 0-3 | 2 | 101.3 | 261 | 235 |
| | 3 | 78.3 | 157 | 186 |
| | 1 | 115.8 | 441 | 385 |
| 0-4 | 2 | 102.7 | 232 | 237 |
| | 3 | 75.4 | 151 | 193 |

**TABLE 2 Parameters of METANET model**

| Parameter | Value |
|:---:|:---:|
| $\tau$ | 73.2 |
| $\kappa$ | 42.0 |
| $\omega$ | 0.16 |
| $\alpha_{TTT}$ | 3.0 |
| $\alpha_{TTD}$ | 2.0 |
| $\alpha_{SD}$ | 5.0 |
| $\gamma$ | 0.7 |

In the simulation scenarios constructed in this paper, rainfall intensity varies with time and space. Rainfall intensity is classified according to the rate of precipitation, which depends on the considered time. The amount of rainfall in an hour is a common international standard for classifying rainfall intensity. Light rain refers to the precipitation rate < 2.5 *mm* per hour; Moderate rain refers to the precipitation rate between 2.5 *mm* to 7.6 *mm*; Heavy rain refers to the precipitation rate is > 7.6 *mm* per hour or between 10 *mm* and 50 *mm* per hour (*29-30*). Table 3 shows the rainfall intensity for each road segment during the simulation time period. In addition, the visibility data collected from the hazemeters was used to fit Eq. 10 and the model parameters obtained are shown in Table 4.

**TABLE 3 Rainfall intensity at each road segment during the simulation time period**

| Time period \ Rainfall intensity | No. 0-1 / mm | No. 0-2 / mm | No. 0-3 / mm | No. 0-4 / mm |
|---|---|---|---|---|
| 0 ~ 900 s | 0.0 | 0.0 | 0.0 | 0.0 |
| 900 ~ 1800 s | 0.5 | 0.5 | 1.5 | 2.5 |
| 1800 ~ 2700 s | 0.5 | 1.0 | 2.5 | 3.5 |
| 2700 ~ 3600 s | 1.0 | 1.5 | 4.0 | 6.0 |

**TABLE 4 Parameters of speed-density model model**

| Parameter | Value |
|---|---|
| A | 0.29 |
| B | 0.17 |
| C | -43.76 |

## RESULTS and DISCUSSION
### Optimal Variable Guidance Speed

The rainfall intensity for each road section during the simulation time period is set above, and it increases with time. Segment No. 0-1 corresponds to the light rain scenario, section No. 0-2 corresponds to the moderate rain scenario, section No. 0-3 corresponds to the heavy rain scenario, and section No. 0-4 corresponds to the torrential rain scenario. **Figure 7** illustrates the variation of guidance speed over time for the different lanes of the four road sections. Overall, as the intensity of rainfall increases, the guidance speeds of different lanes on different segments of the road decrease to varying degrees. In sigment No. 0-1, the guided speed of lane 1 starts with a more pronounced decrease, and then is constant and accompanied by a certain increase in the following; the guided speeds of lane 2 and lane 3 do not decrease significantly but fluctuate [**Figure 7(a)**]. This is mainly due to the higher safety speed limit during light rain and the distance of section 1 from the ramp. As the maximum rainfall intensity increases and the rate of rainfall intensity increase rises, the guidance speeds of lane 1 and lane 2 of segments No. 0-2, No. 0-3, and No. 0-4 all decrease significantly, and both are similar in the later stages of the process [**Figure 7(a)**, **Figure 7(b)**, **Figure 7(c)**].

**Figure 7 (a)**

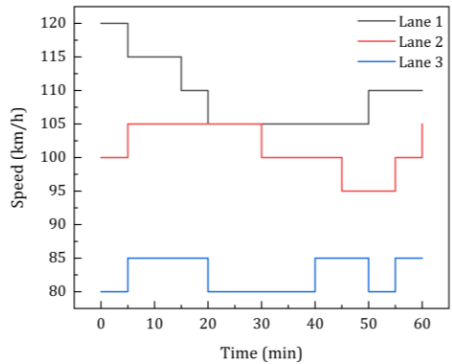

**Figure 7 (b)**

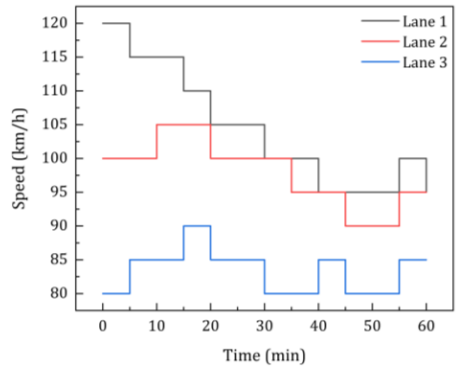

**Figure 7 (c)**

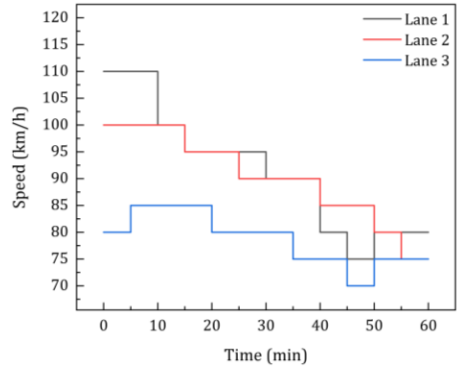

**Figure 7 (d)**

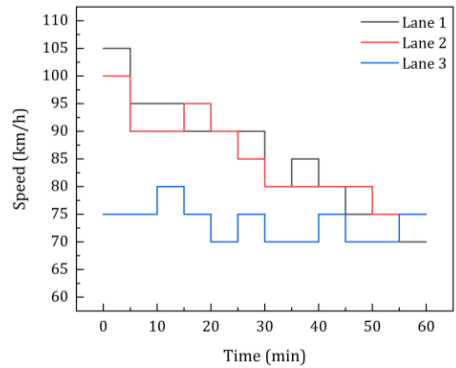

**Figure 7 Guidance speed of segment No. 0-1: (a) Guidance speed of segment No. 0-2: (b) Guidance speed of segment No. 0-3: (c) Guidance speed of segment No. 0-4: (d)**

For lane 3 of segment 4, the guided speed is not only affected by the rainfall intensity but also varies with the distance to the off-ramp. Prior to PDS, guidance speed was only related to control time[**Figure 7(d)**], whereas the guidance speed of PDS varies with time and distance. **Figure 8** illustrates the changes in guidance speeds of PDS. The length of the PDS grows as the intensity of the rainfall increases over time (from 200*m* to 400*m*), due to the smaller acceleration constraint and the longer gradual deceleration process required by the vehicle. At the same time, as the rainfall intensity increases, the initial guidance speed of the PDS is lower (about 75*km/h*), and the final guidance speed is also lower (about 50*km/h*), which well reflects the objective fact that the safe speed limit in rainy days decreases with the increase of rainfall intensity.

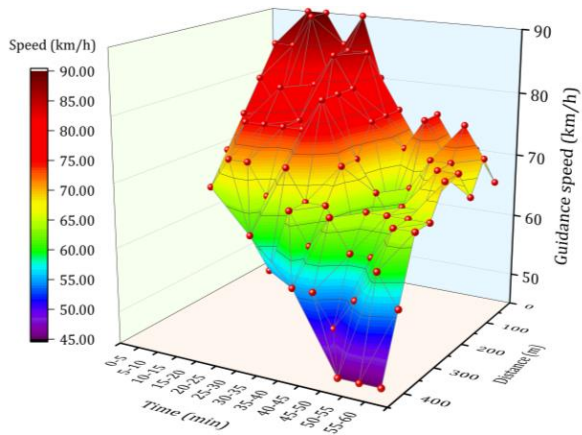

**Figure 8 Variation of PDS guidance speed with time and distance to off-ramp**

**Average Speed**

Distribution of average speeds reflects control effects. Average speed of each segment under the two control modes of fixed speed limit control and joint lane level speed guidance control were compared as shown in Figure 9 (segment 4 lane 3 includes only pre-PDS segment). As the intensity of rainfall increases, the average speed of each road segment decreases to varying degrees under both speed control methods due to reduced visibility. However, the variance of the average speeds of neighboring road segments under the joint lane level speed guidance control is smaller compared to the fixed speed limit control. Under joint lane level speed guidance control, the maximum speed difference between segments 3 and 4 in the same control period is reduced most significantly. It decreases from about 15*km/h* to about 5*km/h*, which indicates a smoother speed change for vehicles approaching the off-ramp. In addition, the average speed of road segments 1 and 2, which have a smaller peak rainfall intensity, is slightly reduced by between 13% and 17%, which reduces capacity to a certain extent, but it is acceptable in terms of traffic safety as it can be articulated with the lower guidance speed term of road segments 3 and 4.

**Figure 9 (a)**

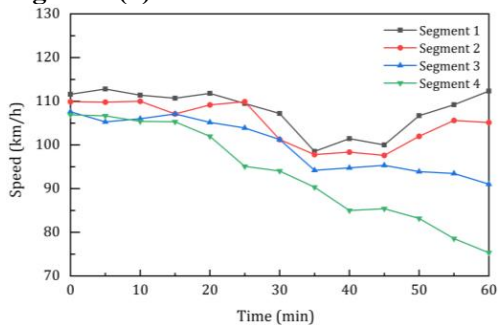

**Figure 9 (b)**

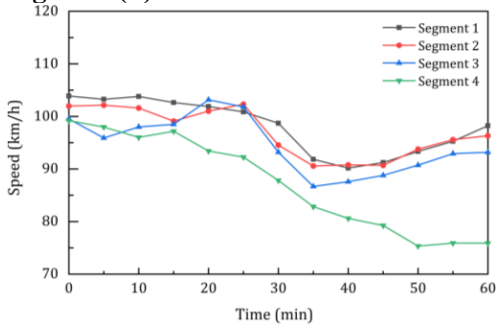

**Figure 9 (c)**

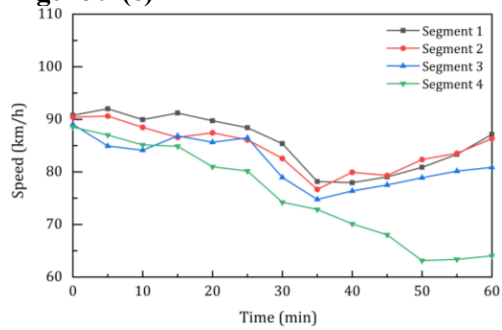

**Figure 9 (d)**

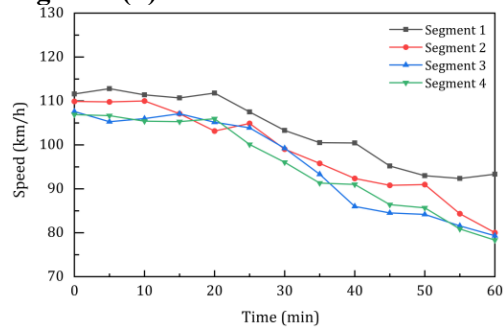

**Figure 9 (e)**

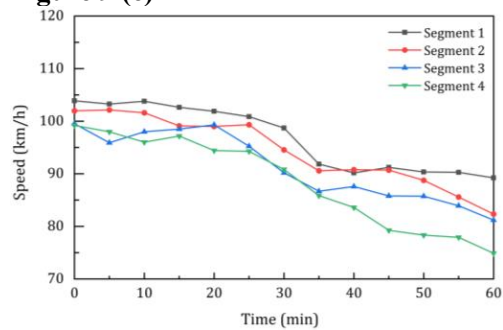

**Figure 9 (f)**

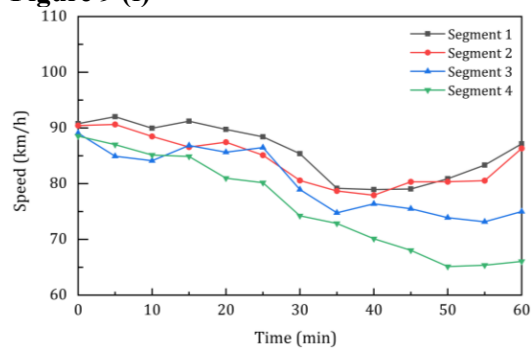

**Figure 9 Average speed of lane 1, 2, 3 under fixed speed limit control: (a), (b), (c) Average speed of lane 1, 2, 3 under joint lane level speed guidance control: (d), (e), (f)**

**Speed of vehicles on the PDS**

For vehicles on the PDS, Figure10 illustrates the spot-speed of vehicles at several observation points (at 0, 100, 200, 300, and 400 meters from the off-ramp) for vehicles. Vehicles are generally able to follow the progressive deceleration guidance on PDS. The distribution of spot-speeds of vehicles at the observation points is in the vicinity of the guidance speeds. About 68.2% of the vehicles had a speed difference of no more than 5*km/h* from the guided speed. At the entrance to the off-ramp, vehicles have mostly slowed to the required safe speeds of off-ramp, About 33.2% of these vehicles were traveling below the guidance speed.

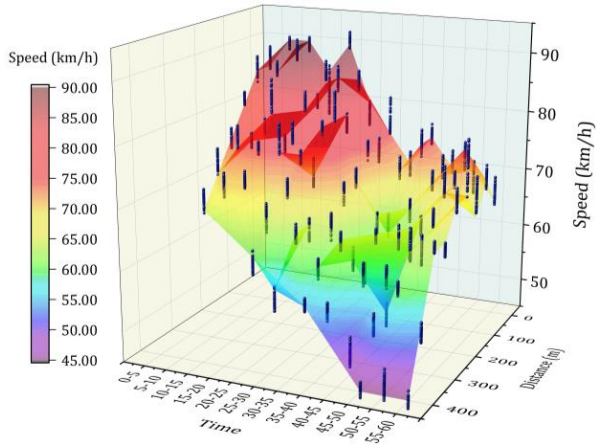

**Figure 10 Point cloud of spot-speed of vehicles on the PDS and guidance speeds on the PDS**

**Sensitivity Analysis**

The control effect is the best when the driver fully complies with the joint lane level speed guidance. However, in reality, due to factors such as driver 's personal driving habits, psychological factors, and communication delays, the driver's compliance rate cannot reach 100 %. Increased noncompliance of driver may lead to weakened performance of joint lane level speed guidance control system. Therefore, it is necessary to explore the impact of different noncompliance rate $\gamma$ of vehicles on the performance of the joint lane level speed guidance control system. The $\gamma$ of vehicles was set from 0 to 1 (e.g. 0, 0.1, 0.2, 0.3, 0.4, 0.5, 0.6, 0.7, 0.8, 0.9, 1.0), and the measures of effectiveness of the joint lane level speed guidance control system were TTT, TTD, SD. Figure 11(a) shows the variation of TTT under various noncompliance rate $\gamma$. With the increase in $\gamma$, the TTT decreased. The simulation results of joint lane level speed guidance control under $\gamma$ = 30%, 50%, and 70% are better than $\gamma$ = 0%, and TTT is reduced by 2.66%, 5.21% and 7.17% respectively. Figure 11(b) shows the variation of TTD under various noncompliance rate $\gamma$. TTD increased significantly with the increase of $\gamma$. Compared with $\gamma$ = 0%, TTD of $\gamma$ = 30%, 50% and 70% increased by 4.90%, 7.07% and 9.48% respectively. Figure 11(c) shows the variation of SD under various noncompliance rate $\gamma$. For the three lanes, SD decreases with the increase of $\gamma$. For example, from $\gamma$ = 0% to $\gamma$ = 100%, the SD of lane 1, lane 2 and lane 3 is improved by 17.59 %, 23.78% and 36.87%.

**Figure 9 (a)**

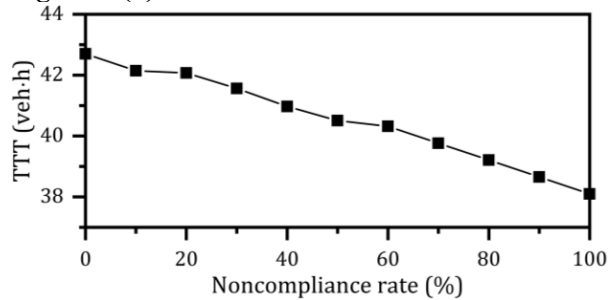

**Figure 9 (b)**

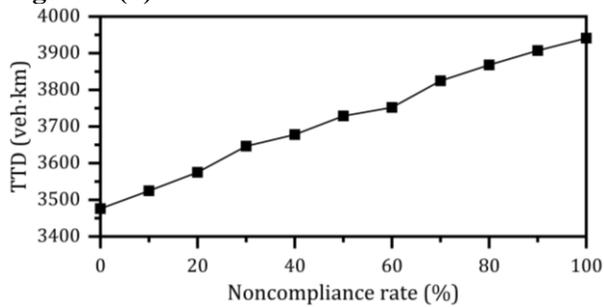

**Figure 9 (c)**

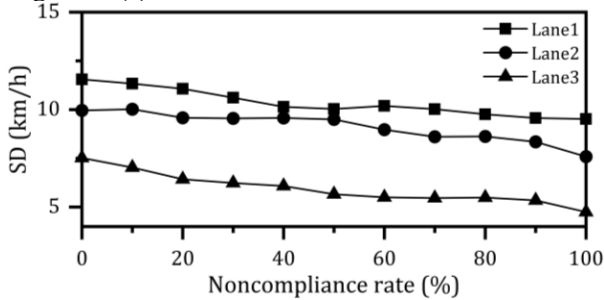

**Figure 11 Performance of joint lane level speed guidance control under various noncompliance rate $\gamma$: (a) TTT with $\gamma$; (b) TTD with $\gamma$; (c) SD with $\gamma$**

## CONCLUSIONS

Based on the existing variable speed limit control technology, this paper proposes a joint speed guidance control strategy for the main line and off-ramp of expressways in rainy weather. There are three innovations: (1) The MATENET model is improved and a lane-level traffic flow prediction model for rainy weather is proposed; (2) A progressive deceleration control method is designed to reduce the speed of the vehicle from the main line speed to the on-ramp safety speed smoothly and safely; (3) A joint speed guidance control model considering the dynamic changes of the safety speed of the main line and ramp in rainy weather is established. This paper selected a real expressway section for simulation, and the developed speed guidance strategy is studied under the background of I2V communication supported by high precision map. Different simulation scenarios are designed considering the intensity of rainfall, and the control performance in different scenarios is compared. The simulation results show that the joint lane-level speed guidance control has more advantages than the existing fixed speed limit mode, which can effectively reduce total travel time and improve traffic efficiency. At the same time, the most important thing is that the speed difference between vehicles and road segments can be significantly reduced under this control method. The safety of driving in rainy weather can be improved. In addition, this study also gives the sensitivity analysis results, with the increase of driver compliance rate, TTT and SD decrease, TTD increase, the better the control effect. In this study, the unsteady evolution process of traffic flow caused by the rapid change of rainfall intensity is not considered. In addition, the joint control strategy that combines speed guidance control and dynamic lane management needs to be further discussed.

**ACKNOWLEDGMENTS**
First of all, This research did not receive any specific grant from funding agencies in the public, commercial, or not-for-profit sectors.

Then, I would like to give my heartfelt thanks to all the people who have ever helped me in this paper. My sincere thanks and appreciations go firstly to my supervisor, Professor Xiaoyun Cheng, whose suggestions and encouragement have given me much insight into these studies. It has been a great privilege and joy to study under her guidance and supervision. I am also extremely grateful to all my friends and classmates who have kindly provided me assistance and companionship in the course of preparing this paper.

Finally, I am really grateful to all those who devote much time to reading this thesis and give me much advice, which will benefit me in my later study.

**AUTHOR CONTRIBUTIONS**
The authors confirm contribution to the paper as follows: study conception and design: Boyao Peng; data collection; code writing: Enkai Li; simulation experiments: Boyao Peng, Lexing Zhang; analysis and interpretation of results: Boyao Peng, Enkai Li; draft manuscript preparation: Boyao Peng. All authors reviewed the results and approved the final version of the manuscript.